\documentclass[preprint]{aa}
\usepackage{graphicx}

\begin{document}

\title{Early star formation in the Galaxy \\
from beryllium and oxygen abundances} 

\author{L. Pasquini\inst{1}, D. Galli\inst{2}, R. G. Gratton\inst{3}
P. Bonifacio\inst{4}, S. Randich\inst{2}, G. Valle\inst{5}}

\institute{European Southern Observatory, Garching bei M\"unchen, Germany
\and
INAF--Osservatorio di Arcetri, Firenze, Italy 
\and
INAF--Osservatorio di Padova, Padova, Italy 
\and
INAF--Osservatorio di Trieste, Trieste, Italy
\and
Dipartimento di Fisica, Universit\`a di Pisa, Italy}

\abstract{We investigate the evolution of the star formation rate in
the early Galaxy using beryllium and oxygen abundances in metal poor
stars. Specifically, we show that stars belonging to two previously
identified kinematical classes (the so-called ``accretion'' and
``dissipative'' populations) are neatly separated in the  [O/Fe] vs.
$\log ({\rm Be/H})$ diagram.  The dissipative population follows the
predictions of our model of Galactic evolution for the thick disk
component, suggesting that the formation of this stellar population
occurred on a timescale significantly longer (by a factor $\sim 5$--10)
than the accretion component. The latter shows a large scatter in the
[O/Fe] vs.  $\log ({\rm Be/H})$ diagram, probably resulting from the
inhomogeneous enrichment in oxygen and iron of the protogalactic gas.
Despite the limitation of the sample, the data suggest that the
combined use of products of spallation reactions (like beryllium) and
elemental ratios of stellar nucleosynthesis products (like [O/Fe]) can
constrain theoretical models for the formation and early evolution of
our Galaxy.
}

\date{Version: 4/18/04}

\authorrunning{L. Pasquini et al.}
\titlerunning{Beryllium and oxygen in the early Galaxy}
\offprints{lpasquin@eso.org}

\maketitle

\section{Introduction}

In a recent paper (Pasquini et al.~2004), we presented the first
detection of beryllium in two turnoff stars of the globular cluster
NGC~6397, and tested the theoretical proposal that Be could be used as
a cosmochronometer for the earliest stages of evolution of the Galaxy.
The rationale behind this suggestion is that Be is produced only by
spallation reactions of Galactic cosmic rays on interstellar medium
nuclei, and if Be was produced by the 
primary process (see e.g. King 2002),  
the evolution of Be was a global process occurring on a
Galactic scale (Beers, Suzuki, \& Yoshii~2000, Suzuki \& Yoshii~2001),
rather than a local process like the production and subsequent ejection
of heavy elements by supernovae (SNe). At any time, the abundance of Be
is thus expected to be characterized by a scatter around the mean value
significantly smaller than that of typical products of stellar
nucleosynthesis like e.g. Fe or O, making Be 
a more reliable
chronometer than [Fe/H] or [O/H]. For example, according to the
stochastic model of Suzuki \& Yoshii~(2001), at a Galactic age of
0.2--0.4~Gyr the spread in $\log({\rm Be/H})$ is about 0.5~dex, whereas
the spread in [Fe/H] is more than twice this value.  This theoretical
prediction was the main justification for the use of Be as a clock for
the early Galaxy in Pasquini et al.~(2004). In this paper we extend this
approach to a sample of halo and thick disk stars, using the Be
abundance as an ``equivalent'' time scale. In particular, we test the 
usefulness of Be as a cosmic clock showing that two
previously identified kinematical classes of low-metallicity stars
likely formed over significantly different time scales. In this spirit, we
attribute most of the scatter in the data, especially at low values of
$\log({\rm Be/H})$, to an intrinsic spread in O and Fe abundances,
rather than Be. This is an idealization, of course, and we expect some
intrinsic scatter to be present in the Be abundance as well.

\section{The sample}

The stars of our sample satisfy two main requirements:  (1) their Be,
Fe and O photospheric abundances are known, and computed in an
homogeneous way, and (2) they belong to one of the two kinematical
classes identified by Gratton et al.~(2003, hereafter G03). The first kinematical
class is composed by a rotating inner population with a galactic
rotation velocity larger than 40~km~s$^{-1}$ and an apogalactic
distance of less than 15~kpc.  This was called by G03
the {\it dissipative collapse} component because it broadly corresponds
to the classical Eggen et al.~(1962) dissipative collapse population,
and includes stars from the classical thick disk and the classical
halo.  The second kinematical class is composed by non rotating or
counter rotating stars, and contains mainly stars of the classical
halo.  It was called the {\it accretion} component, because it can be
roughly identified with the accreted population first proposed by
Searle \& Zinn (1978) to explain the formation of the halo.  These two
components differ not only in their kinematical properties, but also in
their chemical composition (see G03).

The abundances of Be, O, and Fe for the stars of our sample were taken
from Boesgaard et al.~(1999, hereafter B99). To guarantee a homogeneous choice of the
stellar parameters and derived abundances we did not include stars from
other compilations.  We extracted from this sample a metal poor
subsample with [Fe/H] $<-0.5$ to focus only on the earliest phases of
Galactic chemical evolution ($t\la 3$~Gyr).  Also, we eliminated stars
cooler than 5500~K and checked that no evolved star was included in our
sample. Cool stars and giants show Li abundances clearly depleted,
indicating that this element has been burned or diluted (Li and Be are
destroyed in stellar interiors by proton reactions above temperatures
$\sim 2.5$ and $3.5\times 10^6$~K, respectively).

Out of the original 26 stars in the sample of B99
only 20 are left; 4 were eliminated because too metal rich, 1 because
too cool (HD103095), one (BD-13 3442) because no accurate kinematic
parameters were available. We used the abundances obtained with the
King scale, given in Tables 2 and 5 of B99.
Since all data necessary for our work were published in B99, they are not reproduced here.  For a few stars not present
in the sample of G03, the association to the
accretion or the dissipative component was established with the
kinematical parameters derived by Fulbright~(2002).  In summary, twelve
stars belong to the accretion group, eight (HD 19445, 76932, 94028,
134169, 184499, 201889, BD+26 3578, BD +23 3912) to the dissipative
component.

\begin{figure}[t]
\resizebox{9cm}{!}{\includegraphics{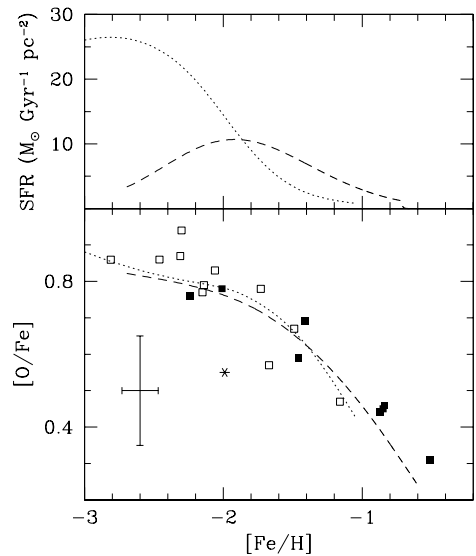}}
\caption{
{\it Upper panel:} the SFR for the halo and the thick
disk ({\it dotted} and {\it dashed curves}, respectively), plotted 
as function of [Fe/H].
{\it Lower panel:} [O/Fe] vs. [Fe/H] diagram for our sample
stars ({\it filled squares}: dissipative component; {\it empty
squares}: accretion component). An {\it asterisk} denotes the position
of the chemically peculiar star HD74000, belonging to the accretion
component. The median absolute abundance error bar is given. 
Note  that the uncertainty in the relative abundances will be smaller.    
The abundance ratios predicted by our model for 
the halo and the thick disk are shown by the 
{\it dotted} and {\it dashed curves},
respectively.} 
\end{figure}

\begin{figure}[t]
\resizebox{9cm}{!}{\includegraphics{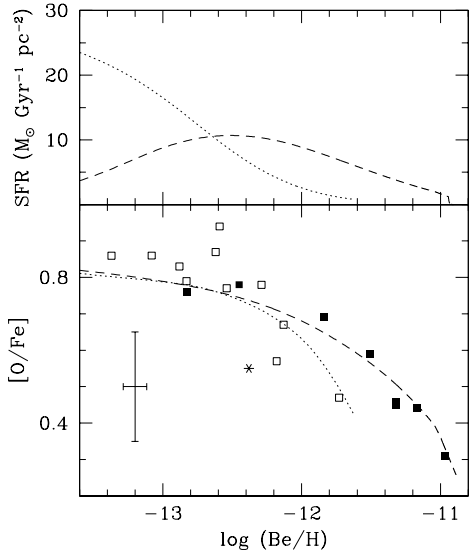}}
\caption{
{\it Upper panel:} the SFR for the halo and the thick disk ({\it
dotted} and {\it dashed curves}, respectively) as function of $\log
({\rm Be/H})$.  {\it Lower panel:} [O/Fe] vs. $\log ({\rm Be/H})$
diagram for our sample stars (same symbols and line types as in
Fig.~1).  Notice the good agreement between the thick disk model and
the dissipative component.
}
\end{figure}

\section{The model of Galactic evolution}

We recall briefly the key features of the model of Galactic evolution
adopted in this work to interpret the observational data summarized in
the previous section.  The model follows the coupled evolution of three
regions in the Galaxy, labelled {\it halo}, {\it thick disk}, and {\it
thin disk} ({\em multizone} treatment), characterized by different star
formation rates and hence exhibiting distinct chemical properties (for
a more detailed discussion see Ferrini et al.~1992, 1994). In the
framework of this model, Valle et al.~(2002) computed the production of
Li, Be, and B nuclei in the Galaxy resulting from spallation reactions
induced by Galactic cosmic rays. According to Valle et al.~(2002), the
production of Be is dominated by spallation of cosmic-ray protons onto
ISM heavy nuclei, mostly O nuclei. At very early times, when the ISM
metallicity is very low, an important contribution to the production of
Be can be provided by spallation of CNO nuclei directly coming from SN
ejecta.  The relative weight of this process depends however on poorly
constrained quantities like the CNO abundance in SN ejecta and the SN
rate at very early times.

As for the stellar production of Fe and O, the heavy elements
considered in this paper, our model follows the nucleosynthesis
prescriptions of Woosley \& Weaver~(1995) and Thielemann, Nomoto \&
Hashimoto~(1996) for SNeII and SNeI, respectively. A detailed
discussion of the evolution of these elements in our model of Galactic
evolution can be found in Travaglio et. al.~(1999). The model
abundances of Fe and O have been converted into abundances relative to
the Sun adopting $\log A({\rm Fe})=7.50$ (Grevesse \& Sauval~1999) and
$\log A({\rm O})=8.69$ (Allende Prieto, Lambert \& Asplund~2001).

\section{Results}

In Fig.~1 we show the [O/Fe] vs. [Fe/H] diagram for our sample stars,
representing the two populations with different symbols (cf. Fig.~3 of
G03), and the abundance ratios predicted by our model
for the halo and the thick disk. The upper panel shows the star
formation rate (SFR) of the model for the two zones, as function of
[Fe/H] (we use the SFR to represent the number of long-living stars
formed at that value of metallicity).  The predictions of our model for
the halo and the thick disk appear to follow a very close evolution in
the [O/Fe] vs. [Fe/H] diagram.  Our sample shows the same
characteristics discussed by G03 for a larger sample,
namely a larger scatter of the accretion component with respect to the
dissipative component, and a slightly higher [O/Fe] ratio for the
latter at fixed metallicity. No clear separation exists between the two
groups, with the obvious exception that the dissipative component
dominates the metal rich tail of the distribution.

The situation is, however, quite different in Fig.~2, showing the
observational data and the model results in the [O/Fe] vs. $\log ({\rm
Be/H})$ diagram (the Be abundance of the model has been normalized to
the solar (meteoritic) value as in Pasquini et al.~2004). The most
interesting feature of Fig.~2 is the significant separation of the halo
and thick disk tracks, especially for high values of $\log({\rm
Be/H})$, and a corresponding separation in the accretion and
dissipation components.  Stars belonging to the dissipative component,
in particular, define in this diagram a sequence characterized by
little or no scatter that closely follows the thick disk track. The
halo track of the model, on the other hand, for high values of
$\log({\rm Be/H})$ defines the upper envelope of the distribution of
stars belonging to the accretion component, but at lower values of
$\log({\rm Be/H})$ the scatter in the data cannot be accounted by our
deterministic model of Galactic evolution. The large spread in the
abundances probably reflects the inhomogeneous enrichment of the halo
gas.

Since oxygen is produced by core-collapse (type-II) SNe, while iron is
also significantly produced by merging-binary (type-I) SNe evolving on
a longer timescale, the oxygen-to-iron ratio is an indicator of the
time variation of the star formation rate at any epoch.  The abundance
of Be, on the other hand, provides a good measure of the time elapsed
after the onset of star formation in the Galaxy, as proposed by Beers
et al.~(2000), Suzuki \& Yoshii (2001) on the basis of stochastic
models of chemical evolution, and empirically demonstrated by Pasquini
et al.~(2004).

Thus, despite the limited sample, the correspondence of the two stellar
populations, identified by G03 on purely kinematic
grounds, with the two components of the model of Galactic evolution,
separated solely on the basis of their different star formation
histories,  supports the idea that the formation of the two populations
took place under significantly different conditions: an inhomogeneous,
rapidly evolving ``halo phase'' for the accretion population, and a
more chemically homogeneous, slowly evolving ``thick disk phase'' for
the dissipative population. As an indication, in our model the SFR in
the halo has a peak around $t\approx 0.06$~Gyr and lasts for about
0.1~Gyr (full width at half-maximum), whereas the SFR in the thick disk
has a peak around $t\approx 0.3$~Gyr, and lasts for about 1~Gyr.

\section{Discussion}

Although the separation of the two stellar populations shown by Fig.~2
suggests a close association between kinematical properties and star
formation history, the identification of the two kinematical groups
with the two components of our Galactic model should be taken with
caution. In fact, the dissipative component identified by G03
also includes halo stars with kinematical properties similar
to those of the thick disk. Also, the use of the [O/Fe] ratio as an
indicator of the variation of the star formation rate is questionable,
as the determination of the O abundance is known to be affected by a
number of uncertainties (see e.g. the discussion in Barbuy et al.
2001).  In principle, a diagram of any $\alpha$ element other than O
vs. $\log({\rm Be/H})$ should present characteristics similar to Fig.~1
and Fig.~2, since all $\alpha$-elements have a common origin in the
pre-supernova stages of massive stars.  We show in Fig.~3 the stars of
our sample for which the [$\alpha$/Fe] ratio was determined by G03, 
as function of [Fe/H] (also from G03) and $\log({\rm Be/H})$. Here,
$\alpha$ is the {\it average} of the abundance ratios of Mg, Si, Ca,
and Ti relative to Fe.  Despite the larger scatter evident in the
``combined'' abundances, and the reduced sample, a significant
separation between the two populations is still discernible, especially
in the [$\alpha$/Fe] vs. $\log({\rm Be/H})$ diagram, lending support to
our conclusions based on the [O/Fe] ratio.

\begin{figure}[t]
\resizebox{9cm}{!}{\includegraphics{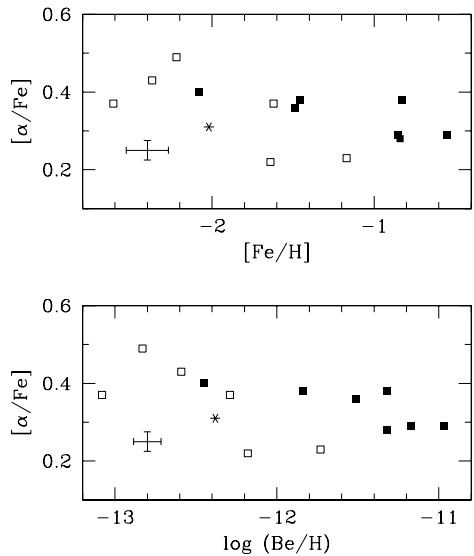}}
\caption{Diagram of [$\alpha$/Fe] 
vs. [Fe/H] and $\log ({\rm Be/H})$, where $\alpha$ is the average of 
the abundances of Mg, Ca, Si, and Ti (same symbols as in Fig.~1 and 2).}
\end{figure}

A feature evident in Figs.~1 and 2 that cannot be reproduced by our
deterministic model of Galactic evolution is the large scatter in
chemical abundances shown by the accretion population.  One should keep
in mind, however, that the amount of intrinsic scatter present in such
a small sample is difficult to assess, since peculiar abundances of
just a few stars may strongly influence the conclusions. For instance,
one of the stars most deviating from the model, HD74000 (shown by an
asterisk in our figures), is known to have a peculiar, N-rich
composition (see e.g. Laird 1985). Some scatter in the accretion
component, however, could be due to spatial inhomogeneities present in
the halo gas.

The halo formation picture emerging from this work favours an accretion
component formed by a short burst plus a component formed in a
long-lasting dissipative process.  This is at odds with the canonical
picture which predicts a very short ($\sim$10$^8$~yr) dissipative
collapse (Eggen et al. 1962) and a rather slow  accretion ($\sim 10
^9$~yr, Searle \& Zinn 1978).  Recent simulations based on $\Lambda$CDM
modelling (Bekki \& Chiba 2001, Samland \& Gerhard 2003) however
suggest that the classical, non rotating halo was formed early by the
disruption of small clumps, and imply a rather long timescale (up to
$\sim 2$~Gyr) for the phase of dissipative collapse.  This picture is
at least in qualitative agreement with our findings. In addition a star
formation process taking place in small clumps may provide a natural
explanation for the $\alpha$/Fe deficiency observed.  Small mass clumps
would in fact not be able to produce many high mass SNae and/or to
retain their ejecta, leading to a net lower Oxygen abundance in the
stars which formed there.  

We should however keep in mind that our analysis is affected by several
uncertainties.  First, our knowledge of the early evolution of galactic
cosmic ray and their confinement is fairly incomplete.  In particular,
the results of our model can strictly be applied only to the solar
neighborhood, whereas it is possible that some halo stars have formed
at very large Galactocentric radii characterized by a lower cosmic ray
flux and therefore a lower production of Be and heavy elements.  In our
sample there is not evidence for a gradient in Be and only one star has
a maximum radius above 20 kpc; the sample is too small to derive any
conclusion in this respect. The absence of an [$\alpha$/Fe] gradient in
the halo (Venn et al. 2004) is in this respect reassuring, but this
possibility should be properly tested before it can be excluded.
Second, our identification of the the outer non rotating component with
the accreted halo, and the rotating inner component with the
dissipative thick disk depends sensitively on the adopted kinematical
criteria.  For example, with the criteria of Venn et al. (2004) only 4
of our stars would be classified as disk objects.

The broad picture outlined in this paper should be confirmed and refined
on the basis of a large, statistically significant sample of stars with
well defined kinematical properties and accurate abundance determinations.
Such a database could, in addition, be used to discriminate different
models of Galactic chemical evolution. Stellar groups proposed to belong
to accretion episodes from external galaxies (Navarro et al. 2004)
could also be studied in a similar fashion.

\begin{acknowledgements}

DG and SR acknowledge financial support by the Italian Ministero 
dell'Istruzione, dell'Universit\`a e della Ricerca through the 
COFIN grant 2002~027319~003. 

\end{acknowledgements}

\end{document}